%% file: CCM_2013_ArXiv.tex
\documentclass[showpacs,pra,tightenlines,12pt]{revtex4-1}
\usepackage{graphicx}
\usepackage{amsmath}
\usepackage{amssymb}
\usepackage{amsthm}
\usepackage{mathrsfs}
\usepackage{amsfonts}
\usepackage[english]{babel}
\usepackage{enumerate}
\usepackage{subfigure}
\usepackage{psfrag}

\newtheorem{defn}{Definition}
\newtheorem*{defn*}{Definition}
\newtheorem{prop}{Property}

\theoremstyle{remark}
\newtheorem{rem}{Remark}
\newtheorem*{rem*}{Remark}

\newtheorem*{examp*}{Example}
\def\Trace{\textrm{Tr}}

\def\ket#1{|#1 \rangle}

\def\qubit#1#2{| #1 \rangle_{#2}}
\def\qubitrho#1#2{| #1 \rangle^{#2} \langle #1|^ {#2}}

\def\NQ{multiqubit }

\def\barq{\bar{q}}

\def\qt{\barq}

\def\prhoK#1{\rho_{#1}}
\def\prhoKA#1{\rho_{A_{#1}}}
\def\prhoKB#1{\rho_{B_{#1}}}
\def\psigmaKA#1{\sigma_{A_{#1}}}
\def\psigmaKB#1{\sigma_{B_{#1}}}

\def\PTr#1{\mathrm{Tr}_{#1}}
\newcommand{\dens}[2]{\left |  #1 \rangle   \langle #2 \right | }

\begin{document}

\title{Cumulative measure of correlation for multipartite quantum states}

\author{Andr\'e L. Fonseca de Oliveira}
\email{fonseca@ort.edu.uy}
\affiliation{Facultad de Ingenier\'ia Bernard Wand-Polaken, Universidad ORT Uruguay, Uruguay.}

\author{Efrain Buksman}
\email{buksman@ort.edu.uy}
\affiliation{Facultad de Ingenier\'ia Bernard Wand-Polaken, Universidad ORT Uruguay, Uruguay.}

\author{Jes\'us Garc\'ia L\'opez de Lacalle}
\email{jglopez@eui.upm.es}
\affiliation{Escuela Universitaria de Inform\'atica, Universidad Polit\'ecnica de Madrid, Spain.}

\date{\today }

\begin{abstract}
The present article proposes a measure of correlation for \NQ mixed states. The measure is defined recursively, accumulating the correlation of the subspaces, making it simple to calculate without the use of regression. Unlike usual measures, the proposed measure is continuous additive and reflects the dimensionality of the state space, allowing to compare states with different dimensions. Examples show that the measure can signal critical points (CP) in the analysis of Quantum Phase Transitions in Heisenberg models.
\end{abstract}


\maketitle

\input{Introduction}

\input{Cumulative_measure}

\input{QPT}

\input{Conclusions}

\appendix

\input{Apend1}

\bibliographystyle{unsrt}
\bibliography{CCM_2013_ArXiv}

\end{document}

%% file: Introduction.tex
\section{Introduction}

Quantum correlations are very significant in quantum information tasks.\cite{Nielsen_2000} For this reason, quantifying correlations is one of the most important problems in quantum information theory.\cite{Bennett_2011}

Many quantum correlation measures have already been proposed: entanglement of formation, geometric entanglement, quantum discord, etc.\cite{Ollivier_2001,Horodecki_2009,Modi_2012} Although the first types of quantum correlation studied were bipartite, today both entanglement and quantum discord have several generalizations for the multipartite case.\cite{Goldbart_2003,Okrasa_2011,Rulli_2011} The relevance of multipartite correlation applies for quantum information processing,\cite{Carvalho_2004,Gallego_2011} for speed up quantum algorithms\cite{BruB_2011} and for the study of many-body systems\cite{Verstraete_2004}. Unfortunately, most of these measures can only be analytically determined  for a few qubits; or require the calculation of nonlinear regressions, which limits the possibilities of studying more complex systems.

In section \ref{Sec_CumulativeCorrelationMeasure} a total correlation measure for \NQ mixed states is defined recursively, accumulating the correlation of the subspaces, and hence, making it simple to calculate without the use of regression. The proposed measure has the property of being a continuous additive measure that reflects the dimensionality of the state. Appendix \ref{App_1} presents proofs for the measure properties.

In Many-body physics analysis, correlation plays a fundamental role in quantum phase transition (QPT). In highly correlated systems, QPT is due to quantum fluctuations.\cite{Sachdev_2007} The behavior of quantum discord and entanglement measures in the analysis of QPT in several Heisenberg spin chain models, shows that correlation measures can be used to detect critical points (CP).\cite{Amico_2008,Rulli_2010,Altintas_2012} As shown in section \ref{Sec_QPT}, the proposed measure serves as an indicator of critical points in QPT, even in low noise environments.

Conclusions and proposed future work are presented in section \ref{Sec_Conclusions}.

%% file: Cumulative_measure.tex
\section{Cumulative correlation measure}
\label{Sec_CumulativeCorrelationMeasure}

\subsection{Definition}
\label{SubSec_CCM_Def}

In this section a definition for a new cumulative measure of correlation for \NQ states is introduced. The proposed correlation computes total (quantum and classical) correlation. 

\begin{defn}
Given a \NQ state $\rho$, the cumulative correlation measure (CCM) is defined as 
\begin{eqnarray}
C(\rho) & = \underset{\{ k \} }{\min} & \left[ 2^{N-2} \ D \left( \rho ,\prhoKA{k} \otimes \prhoKB{k} \right) +  C \left (\prhoKA{k} \right) + C \left( \prhoKB{k} \right) \right] \label{Eqn_Def_cumulativa}
\end{eqnarray}
where $D(\cdot,\cdot)$ denotes a quantum distance, and $k$ is an index for an element $(\prhoKA{k}, \prhoKB{k})$ in the set of all possible bipartitions of the state. 
\end{defn}

The proposed measure starts computing the sum of the distance between state $\rho$ and the product of its reduced matrices ($\prhoKA{k}$ and $\prhoKB{k}$), weighted by a dimensional factor ($2^{N-2}$), accumulating the CCM of each part, $C \left (\prhoKA{k} \right)$ and $C \left( \prhoKB{k} \right)$. The recursion stops for one qubit states ($C(\rho_{1q})=0$). The final result is the minimum among all partitions.

\begin{rem}
In order to achieve the properties listed in the next subsection, the distance $D(\cdot,\cdot)$ used must hold the following features:
\begin{enumerate}
\item \label{Sec_Preliminaries_SubSec_Distance_Prop1} Invariance. $D \left( U \rho U^\dagger,  U \sigma U^\dagger \right) = D ( \rho, \sigma )$, where $U$ is a unitary operator. 
\item \label{Sec_Preliminaries_SubSec_Distance_Prop2} Contractivity. $D \left( \varepsilon ( \rho ),  \varepsilon ( \sigma) \right) \leq D \left( \rho, \sigma \right)$, where $\varepsilon ( \sigma)$ are quantum local operations ($\varepsilon = \varepsilon^1 \otimes \dots \otimes \varepsilon^N $).
\item \label{Sec_Preliminaries_SubSec_Distance_Prop3} Monotonicity. $D \left( \rho_{A}, \sigma_{A} \right) \leq D \left( \rho_{AB}, \sigma_{AB} \right)$. 
\end{enumerate}
\end{rem}

\begin{rem}
In this paper the relative entropy is used
\begin{equation}
\label{Eqn_Def_RelEntropy}
D_{RE} \left( \rho, \sigma \right) \equiv S \left( \rho \| \sigma \right) = \Trace \, \rho \left( \log \rho - \log \sigma \right) 
\end{equation}
as the distance in definition (\ref{Eqn_Def_cumulativa}). Though not a distance, given the symmetric property failure \cite{Audenaert_2005}, it is always used in the order defined in (\ref{Eqn_Def_RelEntropy}). Relative entropy satisfies the aforementioned properties \cite{Nielsen_2000}. Because in the proposed measure $\sigma = \prhoKA{k}  \otimes \prhoKB{k}$, the relative entropy is equal to the mutual information \cite{Henderson_2001}.
\end{rem}

\subsection{Properties of CCM}
\label{SubSec_CCM_Properties}

In the last years some conditions that \NQ correlation measures should satisfy have been discussed\cite{Henderson_2001,Bennett_2011,Brodutch_2012}. Here properties satisfied by CCM are shown, whereas proofs are presented in Appendix \ref{App_1}. 

\begin{prop}
\label{SubSec_CCM_Properties_Prop1}
CCM must be positive or zero.
\begin{equation}
C(\rho) \geq 0.
\end{equation}
\end{prop}

\begin{prop}
\label{SubSec_CCM_Properties_Prop2}
Any product \NQ state $\rho = \rho^1 \otimes \dots \otimes \rho^N$ has no correlation. 
\begin{equation}
C(\rho^1 \otimes \dots \otimes \rho^N) = 0.
\end{equation}
\end{prop}

\begin{prop}
\label{SubSec_CCM_Properties_Prop3}
CCM is invariant under local unitary transformations ($U = U^1 \otimes \dots \otimes U^N$).
\begin{equation}
C(U \rho U^\dagger) = C(\rho).
\end{equation}
\end{prop}

\begin{prop}
\label{SubSec_CCM_Properties_Prop4}
CCM value is not affected in a system $\rho$ increased by locally non-correlated auxiliary subsystems.
\begin{equation}
C(\rho \otimes \sigma) = C(\rho),
\end{equation}
where $\sigma = \sigma^1 \otimes \dots \otimes \sigma^K$.
\end{prop}

\begin{prop}
\label{SubSec_CCM_Properties_Prop5}
Quatum local operations do not increase CCM.
\begin{equation}
C \left( \varepsilon (\rho) \right) \leq C \left( \rho \right),
\end{equation}
where $\varepsilon = \varepsilon^1 \otimes \dots \otimes \varepsilon^N $.
\end{prop}

The above properties are the commonly accepted for a good correlation measure \cite{Zhou_2006}. Besides these, CCM also has the following properties.

\begin{prop}
\label{SubSec_CCM_Properties_Prop6}
CCM is an additive measure.
\begin{equation}
C \left( \phi \otimes \varphi \right) = C \left( \rho \right) + C \left( \varphi \right).
\end{equation}
\end{prop}

\begin{prop}
\label{SubSec_CCM_Properties_Prop7}
CCM reflects the dimensionality of the state space. Two conditions express this:

\begin{enumerate}
\item Maximum correlation measure is non-decreasing with dimensionality.
\begin{equation}
\underset{\{ \mathcal{H_M} \} }{\max} C ( \rho^M ) \leq \underset{\{ \mathcal{H_N} \} }{\max} C ( \rho^N ),
\end{equation}
where $M$ and $N$ are the dimension of the states, and $M \leq N$.

\item For GHZ states \cite{Greenberger_1989} the correlation measure increases with dimension.
\begin{equation}
C \left( \rho_{GHZ_{M}} \right) < C \left( \rho_{GHZ_{N}} \right)
\end{equation}
where $M \leq N$.

\end{enumerate}
\end{prop}

Besides these properties is important to note that the proposed measure is continuous.

\subsection{CCM for GHZ states}
\label{SubSec_CCM_GHZ}

Using decimal notation to simplify the description of canonical base vectors (e.g. $\ket{101} = \ket{5}$), $N$ qubits GHZ states can be described as
\begin{eqnarray}
\ket{GHZ}_{N} & = & \frac{\sqrt{2}}{2} \left( \ket{0 \dots 0} + \ket{1 \dots 1} \right)
= \frac{\sqrt{2}}{2} \left( \ket{0} + \ket{2^N-1} \right), \\
\rho_{GHZ} & = & \dens{GHZ}{GHZ}_{N}.
\end{eqnarray}

For GHZ states, all the bipartite reduced matrices are of type
\begin{eqnarray}
\prhoKA{k} & = & \frac{1}{2} \left( \dens{0}{0} + \dens{2^{N_{Ak}}-1}{2^{N_{Ak}}-1} \right), \label{Eqn_ReducedGHZ} \\
\prhoKB{k} & = & \frac{1}{2} \left( \dens{0}{0} + \dens{2^{N_{Bk}}-1}{2^{N_{Bk}}-1} \right),
\end{eqnarray}
where $N  =  N_{Ak} + N_{Bk}$. The product state in equation (\ref{Eqn_Def_cumulativa}) has the form
\begin{eqnarray}
\prhoKA{k} \otimes \prhoKB{k} & = & \frac{1}{4} \left( \dens{0}{0} + \dens{2^N-1}{2^N-1} + \dots \right. \nonumber \\
& & \left. \dens{m}{m} + \dens{2^N-1 - m}{2^N-1 - m}\right),
\end{eqnarray}
where $m \in \{1, \dots, 2^{N-1}-1 \}$. Essentially, they are all the same state (except for a change in the qubits). Therefore, all the distances between $\rho_{GHZ}$ and $\prhoKA{k} \otimes \prhoKB{k}$ are equal, and are considered normalized (independent of the number of qubits $N$), i.e.
\begin{eqnarray}
D_{GHZ} & \equiv & D \left( \rho_{GHZ} , \prhoKA{k} \otimes \prhoKB{k} \right) = 1, \nonumber \\
\nonumber \\
\Rightarrow C \left( \rho_{GHZ} \right) & = & 2^{N-2}  D_{GHZ} + 
\underset{\{ k \} } {\min} \left[ C \left( \prhoKA{k} \right) + C\left( \prhoKB{k} \right) \right] \nonumber \\
& = & 2^{N-2} + \underset{\{ k \} } {\min} \left[ C \left( \prhoKA{k} \right) + C\left( \prhoKB{k} \right) \right].
\end{eqnarray}

The reduced matrices of $\prhoKA{k}$ are of the same type as (\ref{Eqn_ReducedGHZ}) (as are $\prhoKB{k}$). Then, regardless of the dimension, the normalized distances between A and any of the Kronecker products of its reduced matrices are all the same, denoted here as $d$.

Considering that the minimization for all partitions occurs (due to a power of 2 factor) for partitions with $N/2$ qubits ($N$ even), or with $(N+1)/2$ and $(N-1)/2$ qubits ($N$ odd), the values of CCM for GHZ states can be computed with the following algorithm:
\begin{eqnarray}
F(x, 2) & = & x, \nonumber \\
F(x, 3) & = & 2x + d, \nonumber \\
N \geq 4 \, (\mathrm{even}) : F(x, N) & = & 2^{N-2} x + 2 F(d, \frac{N}{2}), \nonumber \\
N \geq 5 \, (\mathrm{odd}) : F(x, N) & = & 2^{N-2} x + F(d, \frac{N-1}{2})+ F(d, \frac{N+1}{2}), \nonumber \\
C(\rho_{{GHZ}_N}) & = & F(1, N),
\end{eqnarray}
where $\rho_{{GHZ}_N}$ is an $N$ qubit GHZ state. When the distance used is the relative entropy ($D_{RE}$), the constant $d$ equals $1/2$. Table \ref{Tab_GHZ_States} shows the proposed measure for the first nine GHZ states. 

\begin{table}[ph]   %Table~1 
\caption{CCM values for GHZ states GHZ.}
{\begin{tabular}{@{}ccc@{}} \hline 
\\[-1.8ex] 
Number of qubits & CCM & CCM with $D_{RE}$ \\[0.8ex] 
\hline \\[-1.8ex] 
 2 & 1 & 1 \\
 3 & 2 + d & 2.5 \\
 4 & 4 + 2d & 5 \\
 5 & 8 + 4d & 10 \\
 6 & 16 + 6d & 19 \\
 7 & 32 + 9d & 36.5 \\
 8 & 64 + 12d & 70 \\
 9 & 128 + 18d & 137 \\
 10 & 256 + 24d & 268 \\[0.8ex] 
\hline \\[-1.8ex] 
\end{tabular}}
\label{Tab_GHZ_States}
\end{table}

For example, in $\mathcal{H}_4$ the pure states
\begin{eqnarray}
\prhoK{1} = \qubitrho{\psi_1}{}, \quad \qubit{\psi_1}{} & = & \qubit{{GHZ}_2}{} \otimes \qubit{{GHZ}_2}{} \nonumber \\
& = & \frac{1}{2} \left( \qubit{0000}{} + \qubit{0011}{} + \qubit{1100}{} + \qubit{1111}{}\right), \nonumber \\ \nonumber \\
\prhoK{2} = \qubitrho{\psi_2}{}, \quad \qubit{\psi_2}{} & = & \qubit{{GHZ}_3}{} \otimes \qubit{0}{} \nonumber \\
& = & \frac{\sqrt{2}}{2} \left( \qubit{0000}{} + \qubit{1110}{} \right), \nonumber \\ \nonumber \\
\prhoK{3} = \qubitrho{\psi_3}{}, \quad \qubit{\psi_3}{} & = & \qubit{{GHZ}_4}{} = \frac{\sqrt{2}}{2} \left( \qubit{0000}{} + \qubit{1111}{} \right).
\end{eqnarray}
have as results $C\left( \prhoK{1} \right) = 2$, $C\left( \prhoK{2} \right) = 2.5$ and $C\left( \prhoK{3} \right) = 5$. In the first two states ($\prhoK{1}$ and $\prhoK{2}$) the additive property of the measure is explicit. This example shows the influence of large correlated subspaces in high dimension quantum spaces. The larger the dimension of a correlated subspace, the greater the measure of correlation.

%% file: QPT.tex
\section{CCM and quantum phase transition}
\label{Sec_QPT}

In order to show the benefits of the proposed measure, some examples of strongly correlated spin chains, modeled by Heisenberg Hamiltonians, are studied. The generic spin 1/2 Hamiltonian model \cite{Amico_2008} is given by
\begin{equation}
\label{Eq_QPT_Heisenberg}
H = - \sum_{i=1}^{N} \left( J_x \sigma^x_i \sigma^x_{i+1} + J_y \sigma^y_i \sigma^y_{i+1} + J_z \sigma^z_i \sigma^z_{i+1} + h \sigma^z_i \right),
\end{equation}
where $N$ is the number of spins, $\sigma^{x,y,z}$ are the Pauli 
matrices and the boundary condition $\sigma_1=\sigma_{N+1}$ is satisfied. In this paper only the case of interaction through nearest neighbors is analyzed.

Recently, several articles have reported detection of critical points in QPT models by different types of quantum correlations measures, i.e., entanglement, quantum discord and others \cite{Osterloh_2002,Anfossi_2005,Dillenschneider_2008,Cui_2010,Wilms_2012}. The selected examples, the Transverse Ising chain and the XXZ Model, have the advantage that can be exactly solvable in one dimension, and the critical points are well known \cite{Baxter_1982}. The performance of CCM is studied for these models.  

\subsection{The XXZ Model}
\label{SubSec_QPT_XXZ}

\subsubsection{Detection of critical points}
\label{SubSec_QPT_XXZ_Critical}

The Hamiltonian of the anisotropic XXZ model \cite{Sarandy_2009,Yao_2012} is given by
\begin{equation}
\label{Eq_QPT_XXZ}
H_{XXZ} =-\frac{1}{2} \sum_{i=1}^{N} \left( \sigma^x_i \sigma^x_{i+1}+ \sigma^y_i \sigma^y_{i+1} +\Delta \sigma^z_i \sigma^z_{i+1}  \right),
\end{equation}
where $J_x = J_y = 1/2$ and $h=0$, being $J_z = \Delta / 2$ the anisotropy parameter used in (\ref{Eq_QPT_Heisenberg}).

The model exhibits three phases for the ground state: for $\Delta \rightarrow -\infty$ the chain is fully  antiferromagnetic; for $-1 < \Delta < 1$ there is a gapless phase, and for $\Delta \rightarrow \infty$ has a fully polarized ferromagnet. These phases are separated by two critical points: at $\Delta = -1 $ there is an infinite order QPT, and a first order QPT at $\Delta = 1$.

\begin{figure}[!htb]
\centering
\psfrag{text_x}[][]{$\Delta$}
\psfrag{text_y}[bc]{CCM}
\includegraphics[width=0.75\textwidth,keepaspectratio=true]{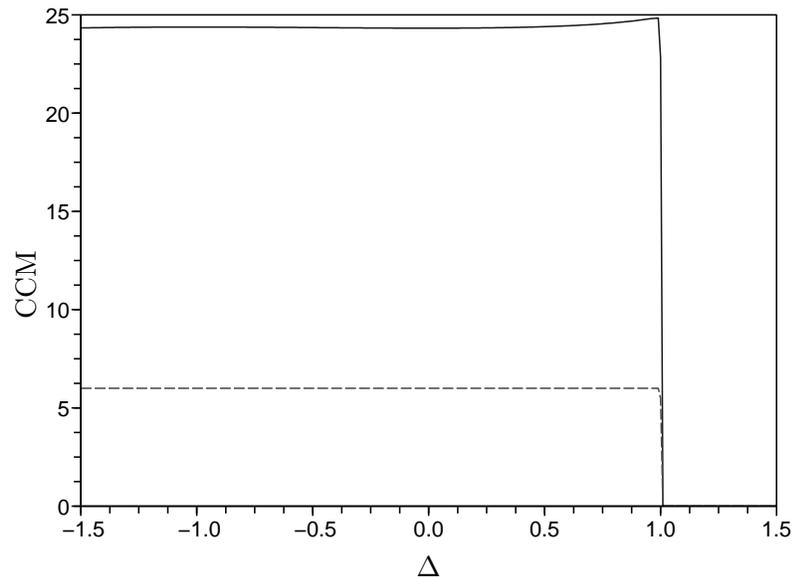}
\caption{Correlation (CCM and $\mathcal{T}_V$) of the grounded state and detection of critical points for XXZ chain of 6 spins. CCM (continuos line) and $\mathcal{T}_V$ (dashed line) for $-1.5 < \Delta < 1.5$. Both measures signaling the first order critial point at $\Delta = 1$.}
\label{Fig_QPT_XXZ_CP_Total}
 \end{figure}

Figure \ref{Fig_QPT_XXZ_CP_Total} shows the results of CCM and the total correlation measure defined in Ref.~\cite{Vedral_2002}, $\mathcal{T}_V$, given by 
\begin{equation}
\label{Eq_QPT_Vedral_measure}
\mathcal{T}_V= S(\rho \| \rho_1 \otimes \rho_2 ... \rho_N).
\end{equation}

As observed, both measures have a discontinuity in $\Delta=1$, signaling a first order critical point.

\begin{figure}[!htb]
\centering
\psfrag{text_x}[][]{$\Delta$}
\psfrag{text_y}[bc]{CCM}
\includegraphics[width=0.75\textwidth,keepaspectratio=true]{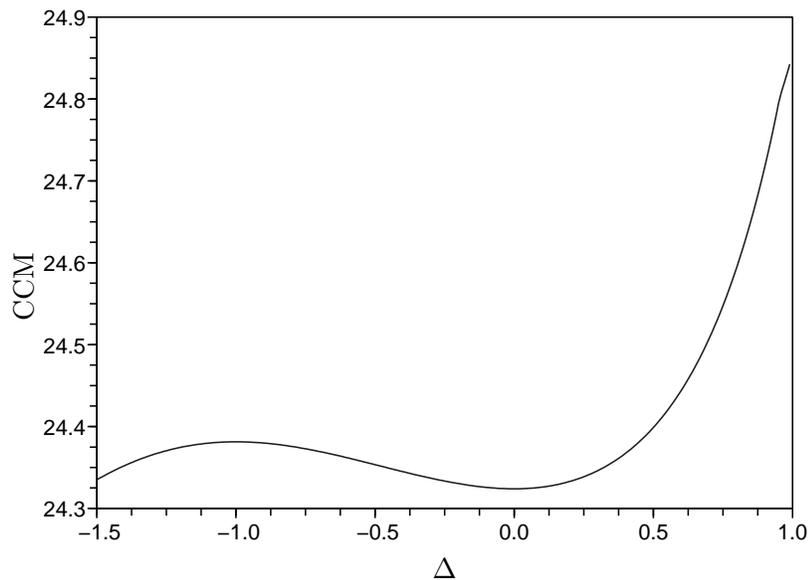}
\caption{Detail ($-1.5 < \Delta < 1$) for CCM in the XXZ model. CCM can detect (maximum) the infinite order critial point at $\Delta = -1$.}
\label{Fig_QPT_XXZ_CP_Detalle}
 \end{figure}

\begin{figure}[!htb]
\centering
\psfrag{text_x}[][]{$\Delta$}
\psfrag{text_y}[bc]{log(CCM)}
\psfrag{text_4q}{4 spins}
\psfrag{text_6q}{6 spins}
\psfrag{text_8q}{8 spins}
\includegraphics[width=0.75\textwidth,keepaspectratio=true]{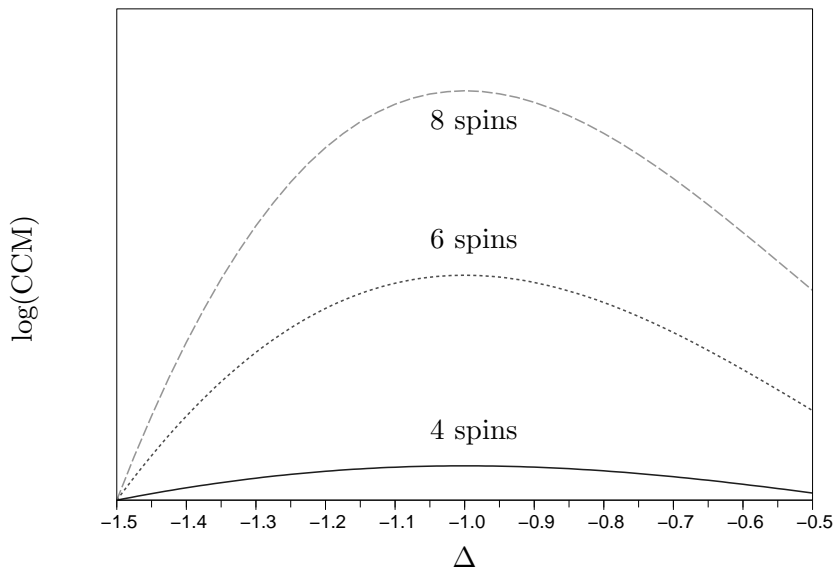}
\caption{CCM for XXZ chains of 4 (solid), 6 (dotted) and 8 (dashed) spins. In order to visualize the differences, the figure shows the logarithm of the measures.}
 \label{Fig_QPT_XXZ_CP_468}
 \end{figure}

Figure \ref{Fig_QPT_XXZ_CP_Detalle} shows the detail for $-1.5 < \Delta < 1$. While for $\mathcal{T}_V$ the value is constant, CCM presents a maximum for $\Delta = -1$  signaling the infinite order critical point. As expected, the maximum value grows with the number of spins in the chain (Figure \ref{Fig_QPT_XXZ_CP_468}).

\subsubsection{Double chain QPT}
\label{SubSec_QPT_Doble}

Unlike genuine measures \cite{Bennett_2011,Giorgi_2011}, CCM can signal the partial phase transition in a system. Consider the double chain Hamiltonian given by
\begin{eqnarray}
\label{Eq_QPT_DXXZ}
H_{DXXZ} = -\frac{1}{2} \sum_{i=1}^{N} & & \left( \sigma^x_i \sigma^x_{i+1}+\tau^x_i \tau^x_{i+1}+ \sigma^y_i \sigma^y_{i+1} + \tau^y_i \tau^y_{i+1} + \cdots \right. \nonumber \\
& & \left. + \Delta \sigma^z_i \sigma^z_{i+1}+\lambda \tau^z_i \tau^z_{i+1} \right),
\end{eqnarray}
where $\sigma^{x,y,z}$ and $\tau^{x,y,z}$ are the Pauli matrices of the two chains. This Hamiltonian represents a double independent XXZ model.

\begin{figure}[!htb]
\centering
\psfrag{text_x}[][]{$\Delta$}
\psfrag{text_y}[][]{$\lambda$}
\psfrag{text_z}[bc]{CCM}
\includegraphics[width=0.75\textwidth,keepaspectratio=true]{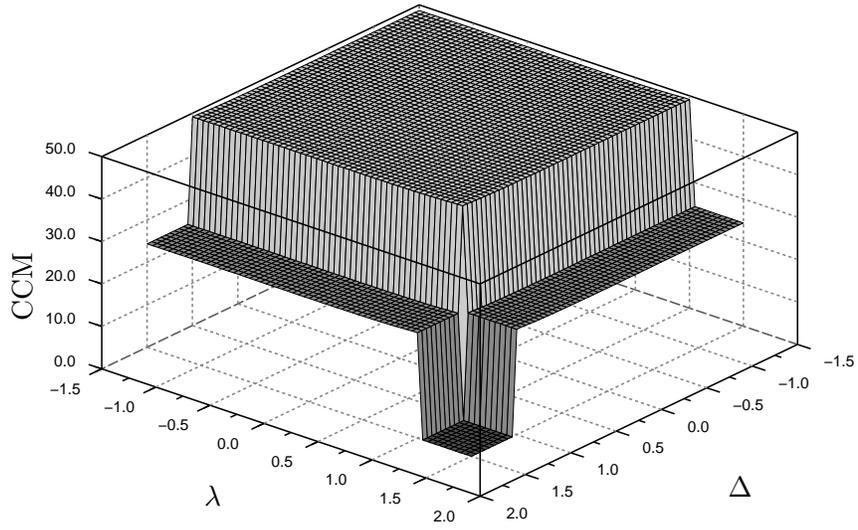}
\caption{Double chain QPT. Independent and simultaneous first order critical points signaling.}
\label{Fig_QPT_XXZ_DB_Total}
 \end{figure}

\begin{figure}[!htb]
\centering
\psfrag{text_x}[][]{$\Delta$}
\psfrag{text_y}[][]{$\lambda$}
\psfrag{text_z}[bc]{CCM}
\includegraphics[width=0.75\textwidth,keepaspectratio=true]{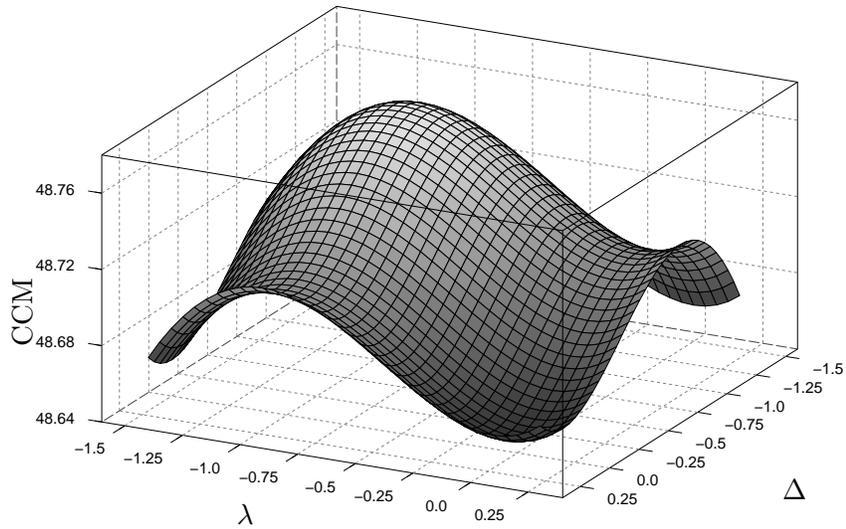}
\caption{Double chain QPT. Detail for the infinity order double critial point at $\Delta = -1$ and $\lambda = -1$.}
\label{Fig_QPT_XXZ_DB_Detalle}
 \end{figure}

The results in figures \ref{Fig_QPT_XXZ_DB_Total} and \ref{Fig_QPT_XXZ_DB_Detalle} show that CCM can signal the partial phase transition in both subsystems.

\subsubsection{Open dynamics and QPT}
\label{SubSec_QPT_XXZ_Open}

In the last years many articles deal with the behavior of the ground state of a spin chain considering the decoherence generated by the interaction with the environment \cite{Werlang_2010,Guo_2013,Werlang_2013}. Real quantum systems are subject to interactions, because they can not be completely isolated. Then, it is of interest to know the behavior of fundamental states near critical points in the presence of noise caused by decoherence. As studied in Ref.~\cite{Pal_2012}, this effect can be achieved using Krauss operator representation, where the decoherence noise is typified by a set of operators $E_i$ that must hold $\sum_i E_i E_i^\dagger = I$.  Then, decoherence can be expressed as 
\begin{equation}
\label{Eq_QPT_Krauss}
\varepsilon(\rho)= \sum_i E_i \rho E_i^\dagger.
\end{equation}

\begin{figure}[!htb]
\centering
\psfrag{text_x}[][]{$p$}
\psfrag{text_y}[][]{$\Delta$}
\psfrag{text_z}{CCM}
\includegraphics[width=0.75\textwidth,keepaspectratio=true]{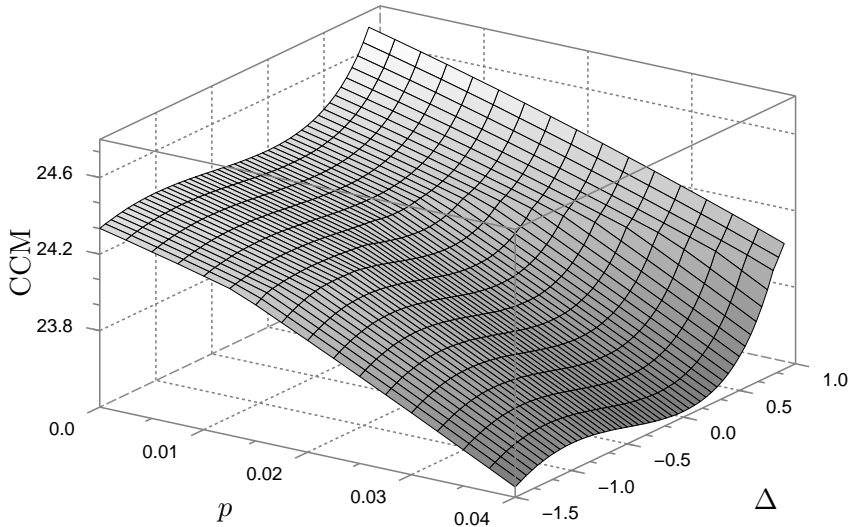}
\caption{CCM evolution of XXZ fundamental states in presence of noise. Evolution of fundamental states near the infinity order critial point.}
\label{Fig_QPT_XXZ_OpenDyn_Total}
 \end{figure}

\begin{figure}[!htb]
\centering
\psfrag{text_x}[][]{$\Delta$}
\psfrag{text_y}{CCM}
\psfrag{text_p_000}[][]{p = 0}
\psfrag{text_p_001}[][]{p = 0.01}
\psfrag{text_p_002}[][]{p = 0.02}
\psfrag{text_p_003}[][]{p = 0.03}
\psfrag{text_p_004}[][]{p = 0.04}
\includegraphics[width=0.75\textwidth,keepaspectratio=true]{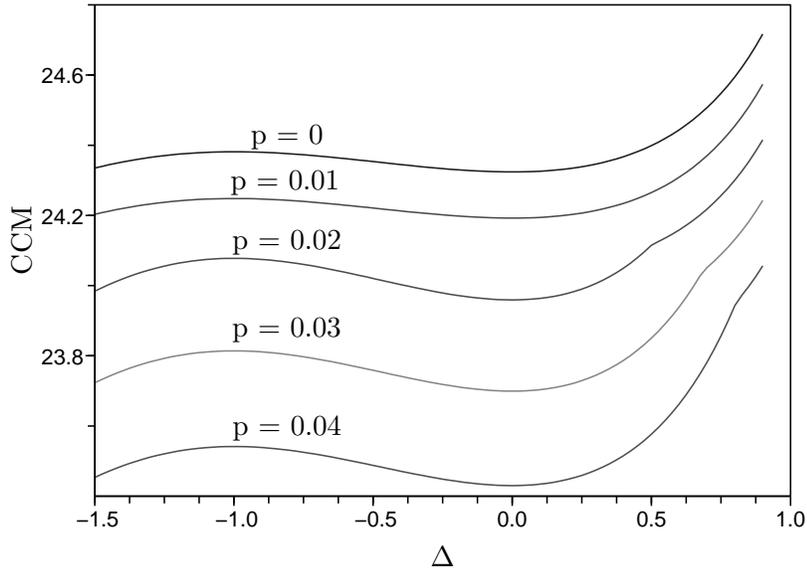}
 \caption{CCM evolution of XXZ fundamental states in presence of noise. Contours for $p \in \{ 0, 0.01, 0.02, 0.03, 0.04 \}$.}
\label{Fig_QPT_XXZ_OpenDyn_Corte}
\end{figure}

This example shows correlation evolution of XXZ fundamental states when amplitude damping noise operators are used, 
\begin{equation}
\label{Eq_QPT_Amplitud_damping}
E_0 = \left(
\begin{matrix}
1 & 0 \\
0 & \sqrt{1-p}
\end{matrix}
 \right) \quad \textrm{and} \quad E_1 = \left(
\begin{matrix}
0 &  0 \\
0 &  \sqrt{p}
\end{matrix}  \right),
\end{equation}
to typify a dissipative interaction with the environment\cite{Nielsen_2000}. In this case, the behavior is similar to the bipartite mutual information showed in Ref.~\cite{Pal_2012}. Figure \ref{Fig_QPT_XXZ_OpenDyn_Total} illustrates the evolution of the ground states near the infinity order critical points ($-1.5 < \Delta < 1$), considering a low noise approach ($0 < p < 0.04$). It is interesting to note that the maximum is more pronounced with some amount of noise (Figure \ref{Fig_QPT_XXZ_OpenDyn_Corte}).

\subsection{Transverse Ising Model}
\label{SubSec_QPT_Ising}

The Ising model has a different universality class of critical point in comparison to the XXZ model. As in Ref.~\cite{Sarandy_2009}, the Hamiltonian of the transverse Ising model is given by (\ref{Eq_QPT_Heisenberg}) using $J_x = 1$, $J_y = J_z = 0$ and $h = \lambda$, resulting in 
\begin{equation}
\label{H_Ising}
H_{Ising} = - \sum_{i=1}^{N} \left( { \sigma^x_i \sigma^x_{i+1} + \lambda \sigma^z_i}\right).
\end{equation}

\begin{figure}[!htb]
\centering
\psfrag{text_x}{$\lambda$}
\psfrag{text_y}[bc]{d(CCM)/d$\lambda$}
\includegraphics[width=0.75\textwidth,keepaspectratio=true]{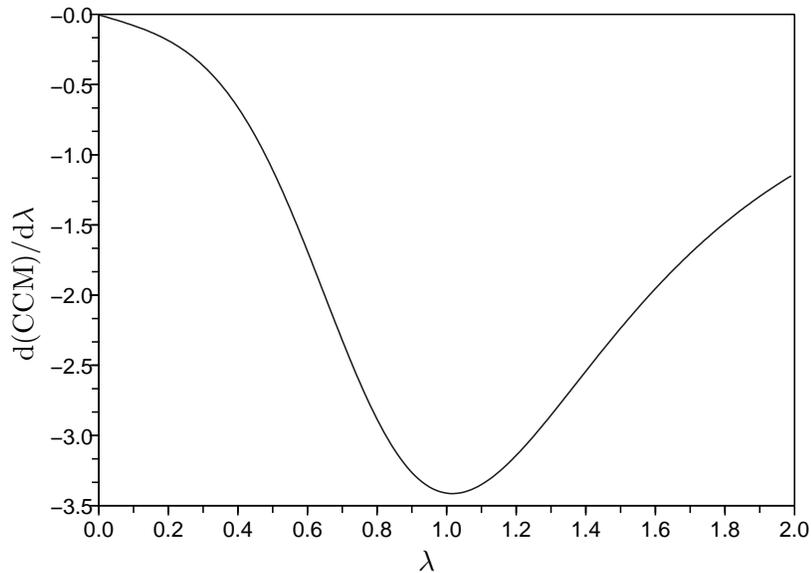}
\caption{Transverse Ising Model. Derivative of CCM near the ground state.}
\label{Fig_QPT_Ising_CCM_DIFF}
\end{figure}

For $\lambda =0 $ all the spins point in the $x$ direction, while for $\lambda \rightarrow \infty $ all point in the $z$ direction.
The ground state presents a critical point at $\lambda = 1$. This is signaled by an inflexion point in the CCM curve (observed in figure \ref{Fig_QPT_Ising_CCM_DIFF} as a minimum in the derivative of CCM curve), similar to other measures of correlation \cite{Dillenschneider_2008,Pal_2012,Auerbach_1998,Osborne_2002}.

%% file: Conclusions.tex
\section{Conclusions}
\label{Sec_Conclusions}

In summary, this paper proposes a total cumulative measure of correlation, CCM, which meets the expected properties for a \NQ correlation, adding desirable features as additivity and dependence on the dimension of space. 

Some results in the study of QPT have been illustrated. As in the case of quantum correlation (as quantum discord), CCM can be used to signaling critical points. And, although the calculation grows exponentially with the dimension, nonlinear regression methods are not required. Another advantage is that for GHZ states the results can be known algebraically, facilitating comparison between systems different dimensions. 

While in this article relative entropy is used as the distance of the algorithm, the proposal is actually a framework that can be defined according to the chosen distance, creating a relationship between each interpretation of correlation and each type of distance.

As future work, it is interesting to find good definitions for quantum correlation counterparts of this total correlation.

\section*{Acknowledgements}

We thank the partial support from the Technology Development Program (PDT, Ministry of Education and Culture of Uruguay) through grant S/C/BE/55/23.

%% file: Apend1.tex
\section{Proof of CCM properties}
\label{App_1}

This section presents the proofs of CCM properties presented in subsection \ref{SubSec_CCM_Properties}.

\subsection*{Property 1}

\begin{proof}
By distance definition, $D \left( \rho, \sigma \right) \geq 0$ for all $\rho$ and $\sigma$. Then, considering the recursive nature of the measure and that $C ( \rho ) = 0$ for the ultimate decomposition (one qubit), $C ( \rho ) \geq 0$ for all $\rho$. 
\end{proof}

\subsection*{Property 2}

\begin{proof}
This property is a direct consequence of the additivity (Property 6) proved below and that the correlation is zero for one-qubits states. 
\end{proof}

\subsection*{Property 3}

\begin{proof}
Assuming a state $\rho' = U \rho U^\dagger$, where  $U = U^1 \otimes \dots \otimes U^N$is a local unitary transformation, the CCM of the new state is
\begin{equation}
C(\rho') = \underset{\{ k \} }{\min} \left[ 2^{N-2} \ D \left( {\rho'} , \prhoKA{k}'   \otimes \prhoKB{k}' \right) + C \left( \prhoKA{k}' \right) + C \left( \prhoKB{k}' \right) \right]
\end{equation}
being $\prhoKA{k}'$ and $\prhoKB{k}'$ the $k$ bipartition of the transformed state. Using the same partition (and properly rearranging the qubits) and considering $U = U_{A_k} \otimes U_{B_k}$ we have that

\begin{equation}
 \prhoKA{k}' = \mathrm{Tr}_{B_{k}} \left( {\rho'} \right) = U_{A_k} \prhoKA{k} U_{A_k}^\dagger, \quad
\prhoKB{k}' = \mathrm{Tr}_{A_{k}} \left( {\rho'} \right) = U_{B_k} \prhoKB{k} U_{B_k}^\dagger,
\end{equation}
which implies that
\begin{equation}
\prhoKA{k}' \otimes \prhoKA{k}' = U_{A_k} \prhoKA{k} U_{A_k}^\dagger \otimes U_{B_k} \prhoKB{k} U_{B_k}^\dagger = U_k \left( \prhoKA{k} \otimes \prhoKB{k} \right) U_k^\dagger.
 \end{equation}

 So, by the required invariance property of the distance
\begin{displaymath}
D \left( {\rho'} , \prhoKA{k}' \otimes \prhoKB{k}' \right) = D \left( \rho_k , \prhoKA{k} \otimes \prhoKB{k} \right).
\end{displaymath}

By finite induction is easy to see that $C \left( \prhoKA{k}' \right) = C \left( \prhoKA{k} \right)$ and $C \left( \prhoKB{k}' \right) = C \left( \prhoKB{k} \right)$, so CCM is invariant under unitary local operations.

\end{proof}

\subsection*{Property 4}

\begin{proof}
This property is a direct consequence of the properties 2 and 6.
\end{proof}

\subsection*{Property 5}

\begin{proof}
Consider local operations such that $\varepsilon (\rho)$, $\varepsilon = \varepsilon^1 \otimes \dots \otimes \varepsilon^N$. The correlation of a state, resulted of applying a local operation $\varepsilon({\rho})$ is
\begin{eqnarray}
C(\varepsilon (\rho ) ) & = \underset{\{ k \} }{\min} & \left[ 2^{N-2} \ D \left( \varepsilon ( \rho ), \psigmaKA{k} \otimes \psigmaKB{k} \right) + C \left( \psigmaKA{k} \right) + C \left( \psigmaKB{k} \right)   \right] 
\end{eqnarray}
where $\varepsilon = \varepsilon_{A_{k}} \otimes \varepsilon_{B_{k}}$, $\psigmaKA{k} = \PTr{B_k}(\varepsilon(\rho))$ and $\psigmaKB{k} = \PTr{A_k}(\varepsilon(\rho))$.

Using Stokes tensor \cite{Jaeger_2007} is straightforward that $\psigmaKA{k} \otimes \psigmaKB{k} = \varepsilon(\prhoKA{k} \otimes \prhoKB{k})$, where $\prhoKA{k} = \PTr{B_k}(\rho)$ and $\prhoKB{k} = \PTr{A_k}(\rho)$. Then, by contractivity (subsection \ref{SubSec_CCM_Def}), 
\begin{eqnarray}
D \left(\varepsilon(\rho),  \psigmaKA{k} \otimes \psigmaKB{k} \right) & = & D \left(\varepsilon(\rho),  \varepsilon(\prhoKA{k} \otimes \prhoKB{k}) \right) \nonumber \\
& \leq & D(\rho, \prhoKA{k} \otimes \prhoKB{k}) \label{Eqn_Apend_Contratividad}
\end{eqnarray}

Using this, we will prove by induction that the proposed measure is contractive, i.e.,
\begin{equation}
C(\varepsilon(\rho)) \leq C(\rho).
\end{equation}

For two qubit states, the property is a direct consequence of the distance between the state and the product of its reduced matrices.

Suppose now that for any state in a space of dimension less than or equal to $N-1$, the
measure is contractive. Consider a \NQ state $\rho$. With local operations, the correlation measure is
\begin{eqnarray}
C(\varepsilon (\rho ) ) & = \underset{\{ k \} }{\min} & \left[ 2^{N-2} \ D \left( \varepsilon ( \rho ), \psigmaKA{k} \otimes \psigmaKB{k} \right) + 
C \left( \psigmaKA{k} \right) + C \left( \psigmaKB{k} \right)   \right].
\end{eqnarray}

For each partition $k$, by (\ref{Eqn_Apend_Contratividad}) we have that
\begin{equation}
D \left(\varepsilon(\rho),  \psigmaKA{k} \otimes \psigmaKB{k} \right) \leq D \left( \rho, \prhoKA{k} \otimes \prhoKB{k} \right).
\end{equation}

As $\psigmaKA{k}$ and $\psigmaKB{k}$ have less than $N$ qubits,
\begin{eqnarray}
C \left( \psigmaKA{k} \right) & \leq & C \left( \prhoKA{k} \right), \nonumber \\
C \left( \psigmaKB{k} \right) & \leq & C \left( \prhoKB{k} \right).
\end{eqnarray}

Then for all $N$,
\begin{equation}
C(\varepsilon(\rho)) \leq C(\rho).
\end{equation}
\end{proof}

\subsection*{Property 6}

\begin{proof}
Now we prove the additivity of the proposed measure. 

For a two qubits product state the property is straightforward (unique partition),
\begin{equation}
C \left( \rho \right) = C \left( \phi \otimes \varphi \right) = 
D \left( \rho, \phi \otimes \varphi \right) + C(\phi) + C( \varphi) = 0.
\end{equation}

In order to perform an induction let's assume that the property holds for all product states with less then $N$ qubits. Considering all possible partitions we have three cases:

\begin{enumerate}

\item The full state $\rho$ is partitioned in the states $\phi$ and $\varphi$. For this particular partition, $k_0$, as $D \left( \rho, \phi \otimes \varphi \right) = 0$, we have that
\begin{equation}
C_{k_0}(\rho) = C(\phi) + C(\varphi).
\end{equation}

\item The used partition $k$ divides the state $\rho$ such that the first partial trace, $\prhoKA{k}$, includes the full state $\phi$ and some part of $\varphi$ (the same applies for the other partial trace). In this case we have
\begin{eqnarray}
\prhoKA{k} & = & \PTr{B_k} \left( \phi \otimes \varphi \right) = \phi \otimes \varphi_{A'_k}, \nonumber \\
\prhoKB{k} & = & \PTr{A_k} \left( \phi \otimes \varphi \right) = \varphi_{B_k},
\end{eqnarray}

where $A'_k$ and $B_k$ is a partition in the $\mathcal{H}_\varphi$ subspace ($A_k$ and $B_k$ is a partition in the hole $ \mathcal{H}_\varphi \otimes \mathcal{H}_\varphi$ space). Then, as additivity holds for states with less than $N$ quibts,
\begin{eqnarray}
C_k(\rho) & = & 2^{N-2} \ D \left( \rho ,\phi \otimes \varphi_{A'_k} \otimes \varphi_{B_k} \right) + C ( \phi ) + \cdots \nonumber \\
& & + C ( \varphi_{A'_k} ) + C ( \varphi_{B_k} ).
\end{eqnarray}

For state $\varphi$ we have that
\begin{eqnarray}
C(\varphi) & = & \underset{\{ k \} }{\min} \left [ C_k(\varphi)\right],  \quad M < N  \nonumber \\
& = & \underset{\{ k \} }{\min} \left[ 2^{M-2} \ D \left( \varphi ,\varphi_{A'_k} \otimes \varphi_{B_k} \right) + C( \varphi_{A'_k} ) + C (\varphi_{B_k}) \right].
\end{eqnarray}

By the monotonicity property (subsection \ref{SubSec_CCM_Def}),
\begin{equation}
D \left( \varphi ,\varphi_{A'_k} \otimes \varphi_{B_k} \right) \leq D \left( \rho ,\phi \otimes \varphi_{A'_k} \otimes \varphi_{B_k} \right).
\end{equation}

Then, 
\begin{equation}
\Rightarrow C_0(\rho) = C(\varphi) + C ( \phi ) \leq C_k(\varphi) + C ( \phi ) \leq C_k(\rho), \quad \forall k.
\end{equation}

\item The used partition $k$ divides the state $\rho$ such that both substates $\phi$ and  $\varphi$ are partitioned, where $A'_k$ and $B'_k$, and $A''_k$ and $B''_k$ are a partitions in the subspaces $\mathcal{H}_\phi$ and $\mathcal{H}_\varphi$, respectively. In this case
\begin{eqnarray}
\prhoKA{k} & = &\PTr{B_k} \left( \phi \otimes \varphi \right) = \phi_{A'_k} \otimes \varphi_{A''_k} \nonumber \\
\prhoKB{k} & = & \PTr{A_k} \left( \phi \otimes \varphi \right) = \phi_{B'_k} \otimes \varphi_{B''_k}
\end{eqnarray}
and $C_k$ is
\begin{eqnarray}
C_k(\rho) & = & 2^{N-2} \ D \left( \rho , \left( \phi_{A'_k} \otimes \varphi_{A''_k}  \right) \otimes \left( \phi_{B'_k} \otimes \varphi_{B''_k} \right) \right) \nonumber \\
& & + C ( \phi_{q'_k} ) + C ( \varphi_{q''_k} ) + C ( \phi_{\qt'_k} ) +  C ( \varphi_{\qt''_k} ).
\end{eqnarray}

By a similar procedure to that used in (b) we have that $C_0 \leq C_k$, $\forall k$. Then, for a \NQ state the property holds, and by induction the measure is additive.
\end{enumerate}
\end{proof}

\subsection*{Property 7}

\begin{enumerate}
\item Condition 1.
\begin{proof}
This property is straightforward. Suppose that $\rho_M \in \mathcal{H}_M$ is the state with maximum measure. In the space $\mathcal{H}_N$, $N = M + L$, we have the state $\phi_N = \rho_M \otimes \varphi_L$. By property 6, $C(\phi_n) = C(\rho_M) + C(\varphi_L) \geq C(\rho_M)$. Then the maximum correlation in $\mathcal{H}_N$ is greater or equal then $C(\rho_M)$.
\end{proof}
\item Condition 2.
\begin{proof}
Proved in subsection \ref{SubSec_CCM_GHZ}.
\end{proof}
\end{enumerate}

%% file: CCM_2013_ArXiv.bbl
\begin{thebibliography}{10}

\bibitem{Nielsen_2000}
Michael Nielsen and Isaac~L. Chuang.
\newblock {\em Quantum Computation and Quantum information}.
\newblock Cambridge University Press, The Edinburgh Building, Cambridge CB2
  2RU, UK, 2000.

\bibitem{Bennett_2011}
C.H. Bennett, A.~Grudka, M.~Horodecki, P.~Horodecki, and R.~Horodecki.
\newblock Postulates for measures of genuine multipartite correlations.
\newblock {\em Phys. Rev. A}, 83:012312, 2011.

\bibitem{Ollivier_2001}
Harold Ollivier and Wojciech~H. Zurek.
\newblock Quantum discord: A measure of the quantumness of correlations.
\newblock {\em Phys. Rev. Lett.}, 88:017901, 2001.

\bibitem{Horodecki_2009}
Ryszard Horodecki, Pawel Horodecki, Michal Horodecki, and Karol Horodecki.
\newblock Quantum entanglement.
\newblock {\em Review of Modern Physics}, 81:865, 942, 2009.

\bibitem{Modi_2012}
K.~Modi, A.~Brodutch, H.~Cable, T.~Paterek, and V.~Vedral.
\newblock The classical-quantum boundary for correlations: Discord and related
  measures.
\newblock {\em Review of Modern Physics}, 84:1655–1707, 2012.

\bibitem{Goldbart_2003}
Tzu-Chieh Wei and Paul~M. Goldbart.
\newblock Geometric measure of entanglement and applications to bipartite and
  multipartite quantum states.
\newblock {\em Phys. Rev. A}, 68:042307, 2003.

\bibitem{Okrasa_2011}
Malgorzata Okrasa and Zbigniew Walczak.
\newblock Quantum discord and multipartite correlations.
\newblock {\em EPL Europhys Letters}, 80:042302, 2011.

\bibitem{Rulli_2011}
C.~C. Rulli and M.~S. Sarandy.
\newblock Global quantum discord in multipartite systems.
\newblock {\em Phys. Rev. A}, 84:042109, 2011.

\bibitem{Carvalho_2004}
A.~R.~R. Carvalho, F.~Mintert, and A.~Buchleitner.
\newblock Decoherence and multipartite entanglement.
\newblock {\em Phys. Rev. Lett.}, 93:230501, 2004.

\bibitem{Gallego_2011}
R.~Gallego, L.~E. Würflinger, A.~Ac\'in, and M.~Navascu\'es.
\newblock Quantum correlations require multipartite information principles.
\newblock {\em Phys. Rev. Lett.}, 107:210403, 2011.

\bibitem{BruB_2011}
D.~Bru\ss \ and C.~Macchiavello.
\newblock Multipartite entanglement in quantum algorithms.
\newblock {\em Phys. Rev. A}, 83:052313, 2011.

\bibitem{Verstraete_2004}
F.~Verstraete, M.~Popp, and J.~I. Cirac.
\newblock Entanglement versus correlations in spin systems.
\newblock {\em Phys. Rev. Lett.}, 92:027901, 2004.

\bibitem{Sachdev_2007}
S.~Sachdev.
\newblock {\em Quantum phase transitions}.
\newblock John Wiley \& Sons Ltd, 2007.

\bibitem{Amico_2008}
L.~Amico, Rosario Fazio, and Andreas Osterloch~Vlatko Vedral.
\newblock Entanglement in many-body systems.
\newblock {\em Rev. Mod. Phys.}, 80:517-- 576, 2008.

\bibitem{Rulli_2010}
C.~C. Rulli and M.~S. Sarandy.
\newblock Entanglement and local extremes at an infinite-order quantum phase
  transition.
\newblock {\em Phys. Rev. A}, 81:032334, 2010.

\bibitem{Altintas_2012}
F.~Altintas and R.~Eryigit.
\newblock Correlation and nonlocality measures as indicators of quantum phase
  transitions in several critical systems.
\newblock {\em Annals of Physics}, 327:3084--3101, 2012.

\bibitem{Audenaert_2005}
K.M.R. Audenaert and J.~Eisert.
\newblock Continuity bounds on the quantum relative entropy.
\newblock {\em J. Math. Phys.}, 46:102104, 2005.

\bibitem{Henderson_2001}
L~Henderson and V~Vedral.
\newblock Classical, quantum and total correlation.
\newblock {\em J. Phys. A: Math. Gen.}, 34:6899, 2001.

\bibitem{Brodutch_2012}
A.~Brodutch and K.~Modi.
\newblock Criteria for measures of quantum correlations.
\newblock {\em Quantum Information and Computation}, 12:0721, 2012.

\bibitem{Zhou_2006}
D.~L. Zhou, B.~Zeng, Z.~Xu, and L.~You.
\newblock Multiparty correlation measure based on the cumulant.
\newblock {\em Phys. Rev. A}, 74:052110, 2006.

\bibitem{Greenberger_1989}
D.~M. Greenberger, M.~Horne, and A.~Zeilinger.
\newblock {\em Bells theorem, quantum theory and conceptions of the universe}.
\newblock Kluwer, 1989.

\bibitem{Osterloh_2002}
A.~Osterloh, Luigi Amico, G.~Falci, and Rosario Fazio.
\newblock Scaling of entanglement close to a quantum phase transition.
\newblock {\em Nature}, 416:608 610, 2002.

\bibitem{Anfossi_2005}
Alberto Anfossi, Paolo Giorda, Arianna Montorsi, and Fabio Traversa.
\newblock Two-point versus multipartite entanglement in quantum phase
  transitions.
\newblock {\em Phys. Rev. Lett.}, 95:056402, 2005.

\bibitem{Dillenschneider_2008}
Raoul Dillenschneider.
\newblock Quantum discord and quantum phase transition in spin chains.
\newblock {\em Phys. Rev. B 78, 224413 (2008) [7 pages]}, 78:224413, 2008.

\bibitem{Cui_2010}
Jian Cui, Jun-Peng Cao, and Heng Fan.
\newblock Quantum information approach to the quantum phase transition in the
  kitaev honeycomb model.
\newblock {\em Phys. Rev. A}, 82:022319, 2010.

\bibitem{Wilms_2012}
J.~Wilms, J.~Vidal, F.~Verstraete, and S.~Dusuel.
\newblock Finite-temperature mutual information in a simple phase transition.
\newblock {\em J. Stat. Mech.}, page 1023, 2012.

\bibitem{Baxter_1982}
J.~R. Baxter.
\newblock {\em Exactly solved models in statistical mechanics}.
\newblock Academic Press, 1982.

\bibitem{Sarandy_2009}
M.S. Sarandy.
\newblock Classical correlation and quantum discord in critical systems.
\newblock {\em Phys. Rev. A.}, 80:022108, 2009.

\bibitem{Yao_2012}
Yao Yao, Hong-Wei Li, Chun-Mei Zhang, Zhen-Qiang Yin, Wei Chen, Guang-Can Guo,
  and Zheng-Fu Han.
\newblock Performance of various correlation measures in quantum phase
  transitions using the quantum renormalization-group method.
\newblock {\em Phys. Rev. A.}, 86:042102, 2012.

\bibitem{Vedral_2002}
V.~Vedral.
\newblock The role of relative entropy in quantum information theory.
\newblock {\em Rev. Mod. Phys.}, 74:197,234, 2002.

\bibitem{Giorgi_2011}
Gian~Luca Giorgi, Bruno Bellomo, Fernando Galve, and Roberta Zambrini.
\newblock Genuine quantum and classical correlations in multipartite systems.
\newblock {\em Phys. Rev. Lett.}, 107:190501, 2011.
\newblock arXiv:1108.0434v2 [quant-ph].

\bibitem{Werlang_2010}
T.~Werlang, C.~Trippe, G.~A.~P. Ribeiro, and Gustavo Rigolin.
\newblock Quantum correlations in spin chains at finite temperatures and
  quantum phase transitions.
\newblock {\em Phys. Rev. Lett.}, 105:095702, 2010.

\bibitem{Guo_2013}
Jin-Liang Guo and Gui-Lu Long.
\newblock Quantum correlation dynamics of three-qubit system coupled to an xy
  chain.
\newblock {\em Eur. Phys. J. D}, 67:53, 2013.

\bibitem{Werlang_2013}
T.~Werlang, G.~A.~P. Ribeiro, and G.~Rigolin.
\newblock Interplay between quantum phase transition and behavior of quantum
  correlations at finite temperatures.
\newblock {\em International Journal of Modern Physics B}, 27:1345032, 2013.

\bibitem{Pal_2012}
A.K. Pal and I.~Bose.
\newblock Transverse ising model: Markovian evolution of classical and quantum
  correlation under decoherence.
\newblock {\em Eur. Phys. J. B}, 85:36, 2012.

\bibitem{Auerbach_1998}
Assa Auerbach.
\newblock {\em Interacting electrons and quantum magnetism}.
\newblock Springer Verlag, Berlin, 1998.

\bibitem{Osborne_2002}
Tobias~J. Osborne and Michael~A. Nielsen.
\newblock Entanglement in a simple quantum phase transition.
\newblock {\em Phys. Rev. A.}, 66:032110, 2002.

\bibitem{Jaeger_2007}
G.~Jaeger.
\newblock {\em Quantum Information: An Overview}.
\newblock Springer New York, 2007.

\end{thebibliography}
